\documentclass[aip,jcp,reprint,onecolumn,amsmath,amssymb,amsfonts,showkeys]{revtex4-1}

\usepackage{graphicx,bm,microtype,hyperref,xcolor,physics}

\usepackage[version=4]{mhchem}
\newcommand{\mc}{\multicolumn}
\newcommand{\eps}{\varepsilon}
\newcommand{\la}{\lambda}

\newcommand{\sbd}{\mathring{\delta}}
\newcommand{\sbD}{\mathring{\Delta}}
\newcommand{\sbzeta}{\mathring{\zeta}}

\newcommand{\Db}{\check{\Delta}}

\newcommand{\sbY}{\mathring{Y}}

\newcommand{\bAB}{\mathbf{AB}}
\newcommand{\bA}{\mathbf{A}}
\newcommand{\bB}{\mathbf{B}}
\newcommand{\bC}{\mathbf{C}}
\newcommand{\bG}{\mathbf{G}}
\newcommand{\bY}{\mathbf{Y}}
\newcommand{\bZ}{\mathbf{Z}}

\newcommand{\bo}{\mathbf{0}}

\newcommand{\ba}{\bm{a}}
\newcommand{\bb}{\bm{b}}
\newcommand{\br}{\bm{r}}
\newcommand{\hD}{\Hat{D}}

\newcommand{\sket}[1]{| #1 ]}
\newcommand{\sexpval}[1]{[ #1 ]}
\newcommand{\sbraket}[2]{[ #1 | #2 ]}

\newcommand{\shield}{\sigma}
\newcommand{\mfac}{\sexpval{\widehat{1}}}

\newcommand{\cmax}[1]{\langle\widehat{#1}\rangle}
\usepackage{tikz}
\usetikzlibrary{arrows,positioning,shapes.geometric}
\usetikzlibrary{decorations.pathmorphing}

\tikzstyle{level 1}=[level distance=5cm, sibling distance=4cm]
\tikzstyle{level 2}=[level distance=5cm, sibling distance=2cm]
\tikzstyle{level 3}=[level distance=5cm, sibling distance=2cm]
\tikzstyle{end} = [circle, minimum width=4pt,fill, inner sep=0pt]

\usepackage{tikz}
\usetikzlibrary{arrows,positioning,shapes.geometric}
\usetikzlibrary{decorations.pathmorphing}

\tikzset{snake it/.style={
decoration={snake, 
    amplitude = .4mm,
    segment length = 2mm},decorate}
}
    
\begin{document}

\title{Three- and four-electron integrals involving Gaussian geminals:\\
fundamental integrals, upper bounds and recurrence relations}
\author{Giuseppe M. J. Barca}
\affiliation{Research School of Chemistry, Australian National University, ACT 2601, Australia}
\author{Pierre-Fran\c{c}ois Loos}
\email{loos@irsamc.ups-tlse.fr}
\affiliation{Laboratoire de Chimie et Physique Quantiques, Universit\'e de Toulouse, CNRS, UPS, France}
\affiliation{Research School of Chemistry, Australian National University, ACT 2601, Australia}

\begin{abstract}
We report the three main ingredients to calculate three- and four-electron integrals over Gaussian basis functions involving Gaussian geminal operators: fundamental integrals, upper bounds, and recurrence relations.
In particular, we consider the three- and four-electron integrals that may arise in explicitly-correlated F12 methods.
A straightforward method to obtain the fundamental integrals is given.
We derive vertical, transfer and horizontal recurrence relations to build up angular momentum over the centers.
Strong, simple and scaling-consistent upper bounds are also reported.
This latest ingredient allows to compute only the $\order{N^2}$ significant three- and four-electron integrals, avoiding the computation of the very large number of negligible integrals.
\end{abstract}

\maketitle

\section{
\label{sec:intro}
Introduction}
It is well known that highly-accurate wave functions require the fulfilment (or near-fulfilment) of the electron-electron cusp conditions. \cite{Kato51, Kato57, Pack66, Morgan93, Myers91, Tew08, QuasiExact09, ExSpherium10, QR12, Kurokawa13, Kurokawa14, Gruneis17}
For correlated wave functions expanded in terms of products of one-electron Gaussian basis functions, the energy converges as $\order{L^{-3}}$, where $L$ is the maximum angular momentum of the basis set. \cite{Kutzelnigg85}  
This slow convergence can be tracked down to the inadequacy of these products to properly model the Coulomb correlation hole. \cite{Hattig12, Kong12}

In the late 20's, Hylleraas solved this issue for the helium atom by introducing explicitly the interelectronic distance $r_{12} = \abs{\br_1 - \br_2}$ as an additional two-electron basis function. \cite{Hylleraas28, Hylleraas29}
As Kutzelnigg later showed, this leads to a prominent improvement of the energy convergence from $\order{L^{-3}}$ to $\order{L^{-7}}$. \cite{Kutzelnigg85}

Around the same time, Slater, while studying the Rydberg series of helium, \cite{Slater28} suggested a new correlation factor known nowadays as a Slater geminal:
\begin{equation}
	S_{12} = \exp(-\lambda_{12}\, r_{12} ).
\end{equation}
Unfortunately, the increase in mathematical complexity brought by $r_{12}$ or $S_{12}$ has been found to be rapidly computationally overwhelming. 
 
In 1960, Boys \cite{Boys60} and Singer \cite{Singer60} independently proposed to use the Gaussian geminal (GG) correlation factor
\begin{equation}
\label{eq:gg}
	G_{12} = \exp(-\lambda_{12}\, r_{12}^2 ),
\end{equation}
as ``\textit{there are explicit formulas for all of the necessary many-dimensional integrals}". \cite{Boys60}
Interestingly, in the same article, a visionary Boys argued that, even if GGs do not fulfil the electron-electron cusp conditions, they could be used to fit $S_{12}$.

During the following years, variational calculations involving GGs flourished, giving birth to various methods, such as the exponentially-correlated Gaussian method. \cite{RychBook, Bubin2005, Bubin2009, Szalewicz2010} 
However, this method was restricted to fairly small systems as it requires the optimization of a large number of non-linear parameters.
In the 70's, the first MP2 calculations including GGs appeared thanks to the work by Pan and King, \cite{Pan70, Pan72} Adamowicz and Sadlej, \cite{Adamowicz77a, Adamowicz77b, Adamowicz78} and later Szalewicz \textit{et al.} \cite{Szalewicz82, Szalewicz83} 
Even if these methods represented a substantial step forward in terms of computational burden, they still require the optimization of a large number of non-linear parameters.

In 1985, Kutzelnigg derived a first form of the MP2-R12 equations using $r_{12}$ as a correlation factor. \cite{Kutzelnigg85} 
Kutzelnigg's idea, which was more formally described together with Klopper in 1987,\cite{Klopper87} dredged up an old problem: in addition to two-electron integrals (traditional ones and new ones), three-electron and four-electron integrals were required.
At that time, the only way to evaluate them would have been via an expensive one- or two-dimensional Gauss-Legendre quadrature. \cite{Preiskorn85, Clementi89}
Nevertheless, the success of the R12 method was triggered by the decision to avoid the computation of these three- and four-electron integrals through the use of the \textit{resolution of the identity} (RI) approximation. \cite{Kutzelnigg91, Hattig12, Werner07}
In this way, three- and four-electron integrals are approximated as linear combinations of products of two-electron integrals.
Several key developments and improvements of the original MP2-R12 approach have been proposed in the last decade. \cite{MOLPRO2011, Hattig12, Werner07, Kong12}
Of course, the accuracy of the RI approximation relies entirely on the assumption that the auxiliary basis set is sufficiently large, i.e.~$N_\text{RI} \gg N$, where $N$ and $N_\text{RI}$ are the number of basis functions in the primary and auxiliary basis sets, respectively.

In 1996, Persson and Taylor killed two birds with one stone.
Using a pre-optimized GG expansion fitting $r_{12}$ 
\begin{equation}
\label{eq:r12fit}
	r_{12} \approx \sum_{k} a_{k} \qty[1-\exp(-\lambda_{k} r_{12}^{2}) ],
\end{equation}
they avoided the non-linear optimization, and eschewed the RI approximation thanks to the analytical integrability of three- and four-electron integrals over GGs. \cite{Persson96}
They were able to show that a six- or ten-term fit introduces a $0.5$ mE$_\text{h}$ or $20$ $\mu$E$_\text{h}$ error, respectively. \cite{Persson96}
Unfortunately, further improvements were unsuccessful due to the failure of $r_{12}$ in modelling the correct behaviour of the wave function for intermediate and large $r_{12}$. \cite{Tew05, Tew06} 
In fact, Ten-no showed that $S_{12}$ is near-optimal at describing the correlation hole, and that a 10-term GG fit of $S_{12}$ yields very similar results.
This suggests that, albeit not catching the cusp per se, the Coulomb correlation hole can be accurately represented by GGs. \cite{Tenno04a, Tenno07, May04, May05} 

Methods for evaluating many-electron integrals involving GGs have already been developed. 
As mentioned previously, Persson and Taylor \cite{Persson97} derived recurrence relations based on Hermite Gaussians, analogously to the work of McMurchie and Davidson for two-electron integrals. \cite{MD78}
These recurrence relations were implemented by Dahle. \cite{DahleThesis, Dahle2007, Dahle2008}
Saito and Suzuki \cite{Saito01} also proposed an approach based on the work by Obara and Saika. \cite{OS1, OS2}
More recently, a general formulation using Rys polynomials \cite{King76, Dupuis76, Rys83} was published by Komornicki and King. \cite{Komornicki11}
Even if limited to the three-center case, it is worth mentioning that May has also developed recurrence relations for two types of three-electron integrals. \cite{MayThesis} 
These recurrence relations were implemented by Womack using automatically-generated code. \cite{WomackThesis}
Recently, we have developed recurrence relations for three- and four-electron integrals for generic correlation factors. \cite{3ERI1, AQC17}

A major limitation of all these approaches is that they do not include any integral screening.\footnote{Komornicki and King mentioned the crucial importance of an effective integral screening in Ref.~\onlinecite{Komornicki11}.}
Indeed, a remarkable consequence of the short-range nature of the Slater and Gaussian correlation factors is that, even if formally scaling as $\order{N^{6}}$ and $\order{N^{8}}$, there are only $\order{N^{2}}$ \textit{significant} (i.e.~greater than a given threshold) three- and four-electron integrals in a large system. \cite{3ERI1, 3ERI2}
Therefore, it is paramount to devise rigorous upper bounds to avoid computing the large number of negligible integrals.

The present manuscript is organized as follows.
In Sec.~\ref{sec:general}, we discuss Gaussian basis functions, many-electron integrals and the structure of the three- and four-electron operators considered here.
The next three sections contain the main ingredients for the efficient computation of three- and four-electron integrals involving GGs: i) fundamental integrals (FIs) in Sec.~\ref{sec:FI}, ii) upper bounds (UBs) in Sec.~\ref{sec:UB}, and iii) recurrence relations (RRs) in Sec.~\ref{sec:RR}. 
In Sec.~\ref{sec:algorithm}, we give an overall view of our algorithm which is an extension of the late-contraction path of PRISM (see Refs.~\onlinecite{HGP, Gill94b} and references therein). 
Note that the RRs developed in this study differ from the ones reported in our previous studies (Refs.~\onlinecite{3ERI1, 4ERI1}) as they are specifically tailored for the unique factorization properties brought by the association of Gaussian basis functions and GGs.
Atomic units are used throughout.

\begin{figure}
	\includegraphics[width=0.6\linewidth]{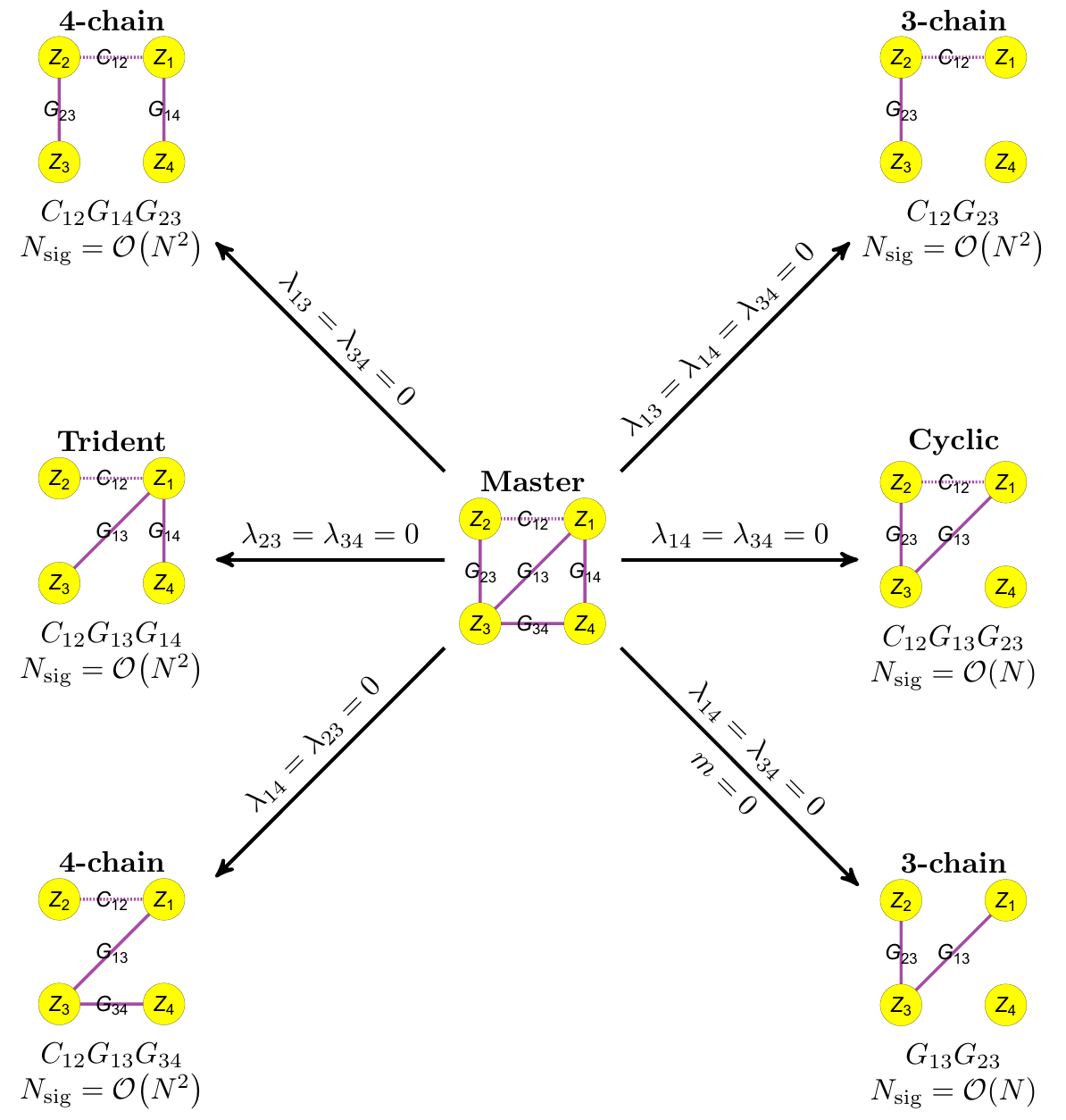}
\caption{
\label{fig:tree}
Diagrammatic representation of the three- and four-electron integrals required in F12 theory.
The number $N_\text{sig}$ of significant integrals in a large system with $N$ CGFs is also reported.}
\end{figure}

\begin{figure*}
	\includegraphics[width=\linewidth]{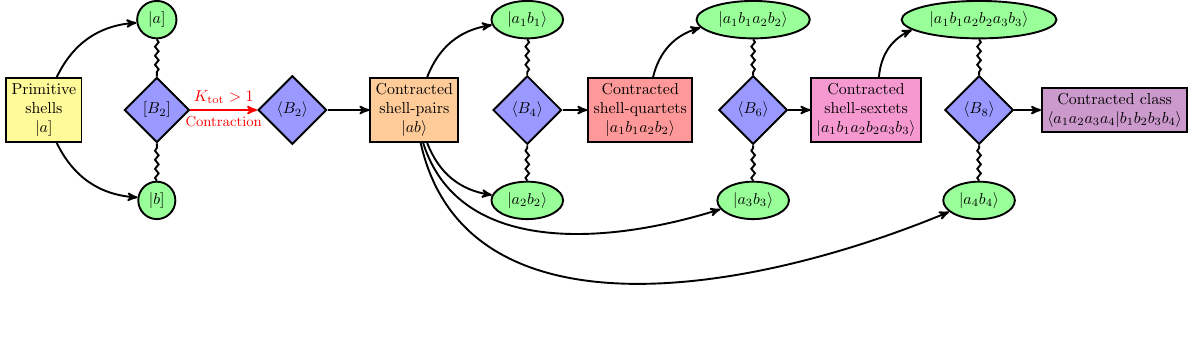}
\caption{
\label{fig:scheme1}
Schematic representation of the screening algorithm used to compute contracted four-electron integrals.}
\end{figure*}

\section{
\label{sec:general}
Generalities}
\subsection{
\label{sec:gaussian}
Gaussian functions}
A primitive Gaussian function (PGF) is specified by an orbital exponent $\alpha$, a center $\bA=(A_x,A_y,A_z)$, and angular momentum $\ba=(a_{x},a_{y},a_{z})$:
\begin{equation}
\label{eq:def1}
	\varphi_{\ba}^{\bA}(\br)  	
	= (x-A_{x})^{a_{x}} (y-A_{y})^{a_{y}} (z-A_{z})^{a_{z}} e^{-\alpha \abs{\br-\bA}^2}.
\end{equation}
A contracted Gaussian function (CGF) is defined as a sum of PGFs
\begin{equation}
\label{eq:def1b}
	\psi_{\ba}^{\bA}(\br)	
	=\sum_{k=1}^{K_a} D_{ak} (x-A_{x})^{a_{x}} (y-A_{y})^{a_{y}} (z-A_{z})^{a_{z}} e^{-\alpha_k \abs{\br-\bA}^2},
\end{equation}
where $K_a$ is the degree of contraction and the $D_{ak}$ are contraction coefficients.
A CGF-pair 
\begin{equation} 
\label{eq:CGTO-pair}
	\ket{\ba \bb} \equiv \psi_{\ba}^{\bA}(\br)\psi_{\bb}^{\bB}(\br) = \sum_{i=1}^{K_a} \sum_{j=1}^{K_b} \sket{\ba \bb}_{ij}
\end{equation}
is a two-fold sum of PGF-pairs $\sket{\ba \bb}=\varphi_{\ba}^{\bA}(\br) \varphi_{\bb}^{\bB}(\br)$.

A primitive shell $\sket{a}$ is a set of PGFs sharing the same total angular momentum $a$, exponent $\alpha$ and center $\bA$. 
Similarly, a contracted shell $\ket{a}$ is a set of CGFs sharing the same PGFs and total angular momentum. 
A contracted shell-pair is the set of CGF-pairs obtained by the tensor product $\ket{a b}=\ket{a} \otimes \ket{b}$.
Similarly, a primitive shell-pair $\sket{a b}=\sket{a} \otimes \sket{b}$ is the set of PGF-pairs.
Finally, primitive and contracted shell-quartets, -sextets and -octets are obtained in an analogous way.
For example, $\sket{a_1 b_1 a_2 b_2} = \sket{a_1 b_1} \otimes \sket{a_2 b_2}$ and $\ket{a_1 a_2 b_1 b_2} = \ket{a_1 b_1} \otimes \ket{a_2 b_2}$.
Note that $\sket{1}$ is a set of three $p$-type PGFs, a $\sket{11} \equiv \sket{pp}$ shell-pair is a set of nine PGF-pairs, and a $\sket{2222} \equiv \sket{dddd}$ shell-quartet is a set of $1,296$ PGF-quartets.

\subsection{
\label{sec:integrals}
Many-electron integrals}
Throughout this paper, we use physicists notations, and we write the integral over a $n$-electron operator $f_{1 \cdots n}$ of CGFs as
\begin{equation}
\begin{split}
\label{eq:def2}
	\braket{\ba_1 \cdots \ba_n}{\bb_1 \cdots \bb_n}
	& \equiv \mel{\ba_1 \cdots \ba_n}{f_{1\cdots n}}{\bb_1 \cdots \bb_n}
	\\
	& = \idotsint 
	\psi_{\ba_1}^{\bA_1}(  \br_{1}) \cdots \psi_{\ba_n}^{\bA_n}(\br_{n})
	\,f_{1\cdots n}\,
	\psi_{\bb_1}^{\bB_1}(  \br_{1}) \cdots \psi_{\bb_n}^{\bB_n}(\br_{n})
	d \br_{1} \cdots d \br_{n}.
\end{split}
\end{equation}
Additionally, square-bracketed integrals denote integrals over PGFs:
\begin{equation}
\label{eq:def2b}
	\sbraket{\ba_1 \cdots \ba_n}{\bb_1 \cdots \bb_n}
	= \idotsint 
	\varphi_{\ba_1}^{\bA_1}( \br_{1}) \cdots \varphi_{\ba_n}^{\bA_n}(\br_{n})
	\,f_{1 \cdots n}\,
	\varphi_{\bb_1}^{\bB_1}( \br_{1}) \cdots \varphi_{\bb_n}^{\bB_n}(\br_{n})
	d \br_{1} \cdots d \br_{n}.
\end{equation}
The FIs (i.e.~the integral in which all $2n$ basis functions are $s$-type PGFs) is defined as $\sexpval{\bo} \equiv \sbraket{\bo \cdots \bo}{\bo \cdots \bo}$ with $\bo=(0,0,0)$.
The Gaussian product rule reduces it from $2n$ to $n$ centers:
\begin{equation}
\label{eq:def4}
	\sexpval{\bo} = 
	\qty( \prod_{i=1}^n S_{i} )
	\idotsint \varphi_{\bo}^{\bZ_1}(\br_{1}) \cdots \varphi_{\bo}^{\bZ_n}(\br_{n})
	\,f_{1 \cdots n}\,d \br_{1} \cdots d \br_{n},
\end{equation}
where 
\begin{subequations}
\begin{align}
	\zeta_i & = \alpha_i + \beta_i,		
	\\
	\bZ_i & = \frac{\alpha_i \bA_i + \beta_i \bB_i}{\zeta_i},		
	\\ 
	S_{i} & = \exp(-\frac{\alpha_i \beta_i}{\zeta_i} \abs{\bA_i\bB_i}^2),
\end{align}
\end{subequations}
and $\bA_i\bB_i = \bA_i - \bB_i$.
We also define the quantity $\bY_{ij} = \bZ_i - \bZ_j$ which will be used later on.

For conciseness, we will adopt a notation in which missing indices represent $s$-type Gaussians.  
For example, $\sexpval{\ba_2\ba_3}$ is a shorthand for $\sbraket{\bo\ba_2\ba_3\bo}{\bo\bo\bo\bo}$.  We will also use unbold indices, e.g. $\sbraket{a_1a_2a_3a_4}{b_1b_2b_3b_4}$ to indicate a complete class of integrals from a shell-octet.

\begingroup
\squeezetable
\begin{table*}
\caption{
\label{tab:bounds}
Primitive bound sets $\sexpval{I_{m}}$ for three- and four-electron integrals and number $N_\text{sig}$ of significant shell-$m$tuplets in a large system with $N$ PGFs.} 
\begin{ruledtabular}
\begin{tabular}{lllclcl}
Integral			&	 type			&	operator				&	\mc{2}{c}{shell-pair level ($m=2$)}	& 	\mc{2}{c}{shell-quartet level ($m=4$)}		
						\\
							\cline{4-5} 						\cline{6-7} 					
				&	&		&	$\sexpval{I_{2}}/\mfac$ 	&	$N_\text{sig}$	&	$\sexpval{I_{4}}$	&	$N_\text{sig}$	
						 \\
						\hline				
three-electron				&	chain	&	
	$C_{12}G_{13}$		&$\left\{\mfac  \sexpval{1}^{12,13} ,\mfac^{13}\sexpval{1}^{12} ,\mfac^{12} \sexpval{1}^{13}  \right\}$		& $\order{N}$ 
						&$\left\{\mfac \sexpval{13}^{12,13}, \mfac^{12} \sexpval{13}^{13} \right\}$					& $\order{N}$  
						\\
						&			&	
	$G_{13}G_{23}$		&$\left\{ \mfac^{23} \sexpval{1}^{13} ,\mfac^{13} \sexpval{1}^{23} ,\mfac \sexpval{1}^{13,23}\right\}$			& $\order{N}$  
						&$\left\{\mfac^{23} \sexpval{13}^{13},\mfac \sexpval{13}^{13,23} \right\}$										& $\order{N}$  
						\\
					&	cyclic	&	
	$C_{12}G_{13}G_{23}$	&$\left\{ \mfac^{23} \sexpval{1}^{12,13} ,\mfac^{13} \sexpval{1}^{12,23} ,\mfac^{12} \sexpval{1}^{13,23}    \right\}$	& $\order{N}$  
						&$\left\{ \mfac \sexpval{13}^{12,13}, \mfac^{12} \sexpval{13}^{13}  \right\}$						& $\order{N}$  
						\\
four-electron				&	chain	&	
	$C_{12}G_{14}G_{23}$	&$\mfac \left\{\mfac^{23} \sexpval{1}^{12,14}, \mfac^{12,14} \sexpval{1}^{23}     \right\}$								& $\order{N}$  
						&$\mfac \left\{ \mfac  \sexpval{14}^{12,14}, \mfac^{12,23} \sexpval{14}^{14} , \mfac^{12} \sexpval{14}^{14}   \right\}$		& $\order{N}$  
						\\
						&			&	
	$C_{12}G_{13}G_{34}$	&$\mfac \left\{\mfac^{34}\sexpval{1}^{12,13}, \sexpval{1}^{34}\mfac^{12,13}    \right\}$									& $\order{N}$  
						&$ \left\{\mfac \mfac^{34} \sexpval{13}^{12,13}, \mfac^{12} \mfac^{34} \sexpval{13}^{13}     \right\}$								& $\order{N}$  
						\\
						&	trident	&	
	$C_{12}G_{13}G_{14}$	&$\mfac \left\{ \sexpval{1}^{12,13,14} \mfac, \sexpval{1} \mfac^{12,13,14}    \right\}$								& $\order{N}$  
						&$  \left\{\mfac \mfac^{14} \sexpval{13}^{12,13}, \mfac^{12}\mfac^{14}\sexpval{13}^{13}     \right\}$								& $\order{N}$  
						\\
						\hline
Integral			&	 type			&	operator				&\mc{2}{c}{shell-sextet level ($m=6$)}				& \mc{2}{c}{shell-octet level ($m=8$)} 
						\\
							\cline{4-5} 						\cline{6-7} 					
					&	&			&$\sexpval{I_{6}}$ & $N_\text{sig}$	&$\sexpval{I_{8}}$ & $N_\text{sig}$ \\
						\hline				
three-electron				&	chain	&	
	$C_{12}G_{13}$		&$\left\{ \sexpval{13}^{12,13} \sexpval{2}, \sexpval{13}^{13}\sexpval{2}^{12} \right\}$	& $\order{N^2}$ 
						&---																& --- 
						\\
						&			&	
	$G_{13}G_{23}$		&$\left\{\sexpval{123}^{13,23} \right\}$									& $\order{N}$  
						&---																& --- 
						\\
					&	cyclic	&	
	$C_{12}G_{13}G_{23}$	&$\left\{\sexpval{123}^{13,23} \right\}$										& $\order{N}$  
						&---																	& --- 
						\\
four-electron				&	chain	&	
	$C_{12}G_{14}G_{23}$	&$\mfac \left\{\sexpval{14}^{12,14} \sexpval{2}^{23},\sexpval{14}^{14} \sexpval{2}^{12,23}    \right\}$		& $\order{N^2}$ 
						&$\left\{\sexpval{14}^{12,14} \sexpval{23}^{23} ,\sexpval{14}^{14} \sexpval{23}^{12,23}    \right\}$	& $\order{N^2}$
						\\
						&			&	
	$C_{12}G_{13}G_{34}$	&$ \left\{ \mfac \sexpval{134}^{12,13,34},  \mfac^{12} \sexpval{134}^{13,34}      \right\}$						& $\order{N}$  
						&$\left\{ \sexpval{134}^{12,13, 34}   \sexpval{2},\sexpval{134}^{13,34}   \sexpval{2}^{12} \right\}$		& $\order{N^2}$ 
						\\
						&	trident	&	
	$C_{12}G_{13}G_{14}$	&$\left\{ \mfac \sexpval{134}^{12,13,14} ,  \mfac^{12}  \sexpval{134}^{13,14}    \right\}$						& $\order{N}$  
						&$\left\{ \sexpval{134}^{12,13,14}   \sexpval{2} ,\sexpval{134}^{13,14}   \sexpval{2}^{12}\right\}$		& $\order{N^2}$ 
						\\
\end{tabular}	
\end{ruledtabular}	
\end{table*}
\endgroup

\subsection{
\label{sec:operators}
Three- and four-electron operators}
In this study, we are particularly interested in the ``master'' four-electron operator $C_{12}G_{13}G_{14}G_{23}G_{34}$ (where $C_{12} = r_{12}^{-1}$ is the Coulomb operator) because the three types of three-electron integrals and the three types of four-electron integrals that can be required in F12 calculations can be easily generated from it (see Fig.~\ref{fig:tree}).
These three types of three-electron integrals are composed by a single type of integrals over the cyclic operator $C_{12}G_{13}G_{23}$, and two types of integrals over the three-electron chain (or 3-chain) operators $C_{12}G_{23}$ and $G_{13}G_{23}$.
F12 calculations may also require three types of four-electron integrals: two types of integrals over the 4-chain operators $C_{12}G_{14}G_{23}$ and $C_{12}G_{13}G_{34}$, as well as one type over the trident operator $C_{12}G_{13}G_{14}$.
Explicitly-correlated methods also requires two-electron integrals. 
However, their computation has been thoroughly studied in the literature. \cite{Kutzelnigg91, Klopper92, Persson97, Klopper02, Manby03, Werner03, Klopper04, Tenno04a, Tenno04b, May05, Manby06, Tenno07, Komornicki11, Tenno12a, Tenno12b, Reine12, Kong12, Hattig12}
Similarly, the nuclear attraction integrals can be easily obtained by taking the large-exponent limit of a $s$-type shell-pair.

Starting with the ``master'' operator $C_{12}G_{13}G_{14}G_{23}G_{34}$, we will show that one can easily obtain all the FIs as well as the RRs required to compute three- and four-electron integrals within F12 calculations.
This is illustrated in Fig.~\ref{fig:tree} where we have used a diagrammatic representation of the operators.
The number $N_\text{sig}$ of significant integrals in a large system with $N$ CGFs is also reported.

\section{
\label{sec:FI}
Fundamental integrals}
Following Persson and Taylor, \cite{Persson96} the $\sexpval{\bo}^{\bm{m}}$ are derived starting from the momentumless integral \eqref{eq:def4} using the following Gaussian integral representation for the Coulomb operator
\begin{equation}
	C_{12} = \frac{2}{\sqrt{\pi}} \int_0^\infty \exp(-u^2 r_{12}^2) du.
\end{equation}
After a lengthy derivation, one can show that the closed-form expression of the FIs is
\begin{equation}
\label{eq:Fund0m}
		\sexpval{\bo}^{m}
		=  \frac{2}{\sqrt{\pi}} \sexpval{\bo}_{G} \sqrt{\frac{\delta_0}{\delta_1-\delta_0}} \qty(\frac{\delta_1}{\delta_1-\delta_0} )^{m} F_m \qty[ \frac{ \delta_1 \qty( Y_1-Y_0 )}{\delta_1-\delta_0} ],
\end{equation}
where $m$ is an auxiliary index, $F_m(t)$ is the generalized Boys function, and
\begin{equation}
\label{eq:Fund0GGGG}
		\sexpval{\bo}_{G}
		= \qty( \prod_{i=1}^4 S_{i} ) \qty( \frac{\pi^4}{\delta_0} )^{3/2} \exp(-Y_0)
\end{equation}
is the FI of the ``pure'' GG operator $G_{13}G_{14}G_{23}G_{34}$ from which one can easily get the FI of the 3-chain operator $G_{13}G_{23}$ by setting $\la_{14} = \la_{34} = 0$.
While the FIs involving a Coulomb operator contain an auxiliary index $m$, the FIs  over ``pure'' GG operators (like $G_{13}G_{23}$) do not, thanks to the factorization properties of GGs. \cite{GG16}

The various quantities required to compute \eqref{eq:Fund0m} are
\begin{equation}
	\bm{\delta}_u
	= \bm{\zeta} + \bm{\la}_u = \bm{\zeta} + \bG + u^2 \bC,
\end{equation}
where
\begin{subequations}
\begin{align}
	\bm{\zeta} & = 
	\begin{pmatrix}
		\zeta_1	&	0		&	0		&	0		\\
		0		&	\zeta_2	&	0		&	0		\\
		0		&	0		&	\zeta_3	&	0		\\
		0		&	0		&	0		&	\zeta_4	\\
	\end{pmatrix},
	\quad
	\bC	= 
	\begin{pmatrix}
		1	&	-1	&	0	&	0	\\
		-1	&	1	&	0	&	0	\\
		0	&	0	&	0	&	0	\\
		0	&	0	&	0	&	0	\\
	\end{pmatrix},
	\\
	\bG & = 
	\begin{pmatrix}
		\la_{13}+\la_{14}		&	0		&	-\la_{13}				&	-\la_{14}			\\
		0					&	\la_{23}	&	-\la_{23}				&	0				\\
		-\la_{13}				&	-\la_{23}	&	\la_{13}+\la_{23}+\la_{34}	&	-\la_{34}				\\
		-\la_{14}				&	0		&	-\la_{34}				&	\la_{14}+\la_{34}	\\
	\end{pmatrix},
\end{align}
\end{subequations}
and
\begin{subequations}
\begin{align}
	 \bm{\Delta}_u & = \bm{\zeta} \cdot \bm{\delta}_u^{-1} \cdot \bm{\zeta},
	&
	\bY^k & = 
	\begin{pmatrix}
		0	&	\bY_{12}^k	&	\bY_{13}^k	&	\bY_{14}^k	\\
		0	&	0			&	0			&	\bY_{24}^k	\\
		0	&	0			&	0			&	\bY_{34}^k	\\
		0	&	0			&	0			&	0			\\
	\end{pmatrix},
	\\
	\delta_u & = \det(\bm{\delta}_u),
	&
	Y_u & =  \Tr( \bm{\Delta}_u \cdot \bY^2).
\end{align}
\end{subequations}
The generalized Boys function $F_m(t)$ in Eq.~\eqref{eq:Fund0m} can be computed efficiently using well-established algorithms. \cite{Gill91, Ishida96, Weiss15}

\section{
\label{sec:UB}
Upper bounds}
In this section, we report UBs for primitive and contracted three- and four-electron integrals.
Our UBs are required to be simple (i.e.~significantly computationally cheaper that the true integral), strong (i.e.~as close as possible to the true integral in the threshold region $10^{-14}-10^{-8}$), and scaling-consistent (i.e.~the number of significant integrals $N_\text{sig}=\mathcal{O}(N_\text{UB})$, where $N_\text{UB}$ is the number of integrals estimated by the UB).
We refer the interested reader to Refs.~\onlinecite{Gill94a, GG16} for additional information about UBs.
A detailed study of these UBs (as well as their overall performances) will be reported in a forthcoming paper. \cite{3ERI2}

Our screening algorithms are based on primitive, $[B_{m}]$, and contracted, $\expval{B_{m}}$, shell-$m$tuplet bounds.   
These are based on shell-$m$tuplet information only: shell-pair ($m=2$), shell-quartet ($m=4$), shell-sextet ($m=6$) and shell-octet ($m=8$). 
Thus, for each category of three- and four-electron integrals, we will report from shell-pair to shell-sextet (or shell-octet) bounds. 

Figure \ref{fig:scheme1} is a schematic representation of the overall screening scheme for contracted four-electron integrals.
First, we use a primitive shell-pair bound $\sexpval{B_2}$ to create a list of significant primitive shell-pairs.
For a given contracted shell-pair, if at least one of its primitive shell-pairs has survived, a contracted shell-pair bound $\expval{B_2}$ is used to decide whether or not this contracted shell-pair is worth keeping.
The second step consists in using a shell-quartet bound $\expval{B_4}$ to create a list of significant contracted shell-quartets by pairing the contracted shell-pairs with themselves.
Then, we combine the significant shell-quartets and shell-pairs, and a shell-sextet bound $\expval{B_6}$ identifies the significant contracted shell-sextets.  
Finally, the shell-sextets are paired with the shell-pairs. 
If the resulting shell-octet quantity is found to be significant, the contracted integral class $\braket{a_1a_2a_3a_4}{b_1b_2b_3b_4}$ must be computed via RRs, as discussed in the next section.
The number of significant shell-$m$tuplets generated at each step is given in Table \ref{tab:bounds}.
As one can see, the size of any shell-$m$tuplet list is, at worst, quadratic in a large system.

During the shell-pair screening, either a contracted or a primitive path is followed depending on the degree of contraction of the integral class $K_\text{tot}=\prod_{i=1}^{n} K_{a_{i}}K_{b_{i}}$.
If $K_\text{tot}>1$, the contracted path is enforced, otherwise the primitive path is followed.
This enables to adopt the more effective primitive bounds for primitive integral classes which are usually associated with medium and high angular momentum PGFs and, therefore, are more expensive to evaluate via RRs.
The scheme for primitive four-electron integrals differs only by the use of primitive bounds instead of contracted ones.
The three-electron integrals screening scheme can be easily deduced from Fig.~\ref{fig:scheme1}.

Note that we bound an entire class of integrals with a single UB.
This is a particularly desirable feature, especially when dealing with three- or four-electron integrals where the size of a class can be extremely large. 
For example, the simple $\sbraket{ppp}{ppp}$ and $\sbraket{pppp}{pppp}$ classes are made of 729 and 4,096 integrals!

\begin{table}
\caption{
\label{tab:fact}
Bounds factors for three- and four-electron integrals.
$\sbzeta_i=(1-\shield_i)\zeta_i$, and $h_i$, $\sbd_u$ and $\sbY$ are given by Eqs.~\eqref{eq:h}, \eqref{eq:sbd} and \eqref{eq:sbY}.
} 
\begin{ruledtabular}
\begin{tabular}{lc}
	Bound factor										&	expression		\\					
	\hline
	$\sexpval{1}^{13,14}$				&	$\displaystyle h_{1} \left(\frac{\pi}{\sbzeta_{1}+\lambda_{13}+\lambda_{14}}\right)^{3/2} $		\\
	$\sexpval{1}^{12,13,14}$				&	$\displaystyle h_{1} \frac{2\pi}{\sbzeta_{1}+\lambda_{13}+\lambda_{14}} $ \\
	$\sexpval{13}^{13}$					&	$\displaystyle h_{1}h_{3}\left( \frac{\pi^2}{\sbd_{0}^{13}} \right)^{3/2} \exp(-\sbY^{{13}})$ \\
	$\sexpval{13}^{12,13}$				&	$\displaystyle \frac{2}{\sqrt{\pi}} \sqrt{\frac{\sbd_{0}^{13}}{\sbd_{1}^{13}-\sbd_{0}^{13}}} \sexpval{13}^{13}$	\\
	$\sexpval{134}^{13,14}$				&	$\displaystyle h_{1}h_{3}h_{4} \left( \frac{\pi^3}{\sbd_{0}^{13,14}} \right)^{3/2} \exp(-\sbY^{{13,14}})$ \\[5pt]
	$\sexpval{134}^{12,13,14}$			&	$\displaystyle \frac{2}{\sqrt{\pi}} \sqrt{\frac{\sbd_{0}^{13,14}}{\sbd_{1}^{13,14}-\sbd_{0}^{13,14}}} \sexpval{134}^{13,14}$	\\
\end{tabular}	
\end{ruledtabular}
\end{table}

\subsection{
\label{sec:pb}
Primitive bounds}
In this section we present UBs for primitive three- and four-electron integrals.
Without loss of generality, we assume that the geminal operators are ordered by decreasing exponent, i.e.~$\la_{13} \le \la_{14} \le \la_{23} \le \la_{34}$.

All the required primitive bounds have the form
\begin{equation}
\label{eq:fineB}
	\sexpval{B_{m}}=
	\begin{cases}
			\max \, \sexpval{I_{m}},	&	m=2,
			\\
			\min \, \sexpval{I_{m}},	&	m>2,
	\end{cases}
\end{equation}
where $m$ is the shell multiplicity. 
The bound sets $\sexpval{I_{m}}$ are reported in Table \ref{tab:bounds}. 
They also require the bound factors in Table \ref{tab:fact}, which are easily computed with the following quantities:
\begin{subequations}
\begin{gather}
	\label{eq:h}
	h = \abs{D_{a} D_{b}} \gamma_{a}\gamma_{b} \sqrt{\frac{a^{a} b^{b}}{e^{a+b}}} \frac{\qty(4 \alpha \beta)^{3/4}}{ \shield^{\frac{a+b}{2}} }
	e^{- \frac{(1-\shield)\abs{\bAB}^{2}}{\alpha^{-1}+\beta^{-1}}},
	\\
	\label{eq:sbd}
	\sbd_{u}^{{13},{14},{23},{34}} = \det(\bm{\sbd}_{u}^{{13},{14},{23},{34}}),
	\\
	\label{eq:sbY}
	\sbY^{{13},{14},{23},{34}} =  \Tr(  \bm{\sbD}_{0}^{{13},{14},{23},{34}} \cdot \bY^2),
\end{gather}
\end{subequations}
where
\begin{subequations}
\begin{gather}
	\gamma_{a}= \prod^{2}_{i=0} \qty[ \Gamma\qty(\left\lfloor\frac{a+i}{3}\right\rfloor+\frac{1}{2} )]^{-1/2},
	\\
	\bm{\sbd}_{u}^{{13},{14},{23},{34}}
	= \bm{\sbzeta} + \bm{\la}_{u} 
	= 
	\begin{pmatrix}
		\sbzeta_{1}	&	0		&	0		&	0		\\
		0		&	\sbzeta_{2}	&	0		&	0		\\
		0		&	0		&	\sbzeta_{3}	&	0		\\
		0		&	0		&	0		&	\sbzeta_{4}	\\
	\end{pmatrix}
	+
	\bm{\la}_{u},
	\\
	 \bm{\sbD}_{u}^{{13},{14},{23},{34}} = \bm{\sbzeta} \cdot \left(\bm{\sbd}_{u}^{{13},{14},{23},{34}} \right)^{-1} \cdot \bm{\sbzeta},
\end{gather}
\end{subequations}
$\Gamma(x)$ and $\lfloor x \rfloor$ are the Gamma and floor functions respectively, \cite{NISTbook} and $\sbzeta_i= (1-\shield_i)\zeta_i$.
We point out that bound factors $\sexpval{1}^{12}$ or $\sexpval{1}^{12,13}$ reported in Table \ref{tab:fact} can be obtained from $\sexpval{1}^{12,13,14}$  by setting $\la_{13}=\la_{14}=0$ or $\la_{14}=0$, respectively.
The parameter $\sigma_1$ is obtained by solving the quadratic equation
\begin{equation}
\label{eq:pm2}
	\pdv{\sexpval{1}^{{13},{14},{23},{34}}}{\shield_{1}} = 0.
\end{equation}

Factors of the kind $\sexpval{\widehat{1}}^{{13},{14},{23},{34}}$ are also required for the bounds in Table \ref{tab:bounds}.
They are defined as the largest factor $\sexpval{1}^{{13},{14},{23},{34}}$ within a given system and basis set, and can be pre-computed and stored with the remaining basis set information. 

\begin{figure}
	\includegraphics[width=0.6\linewidth]{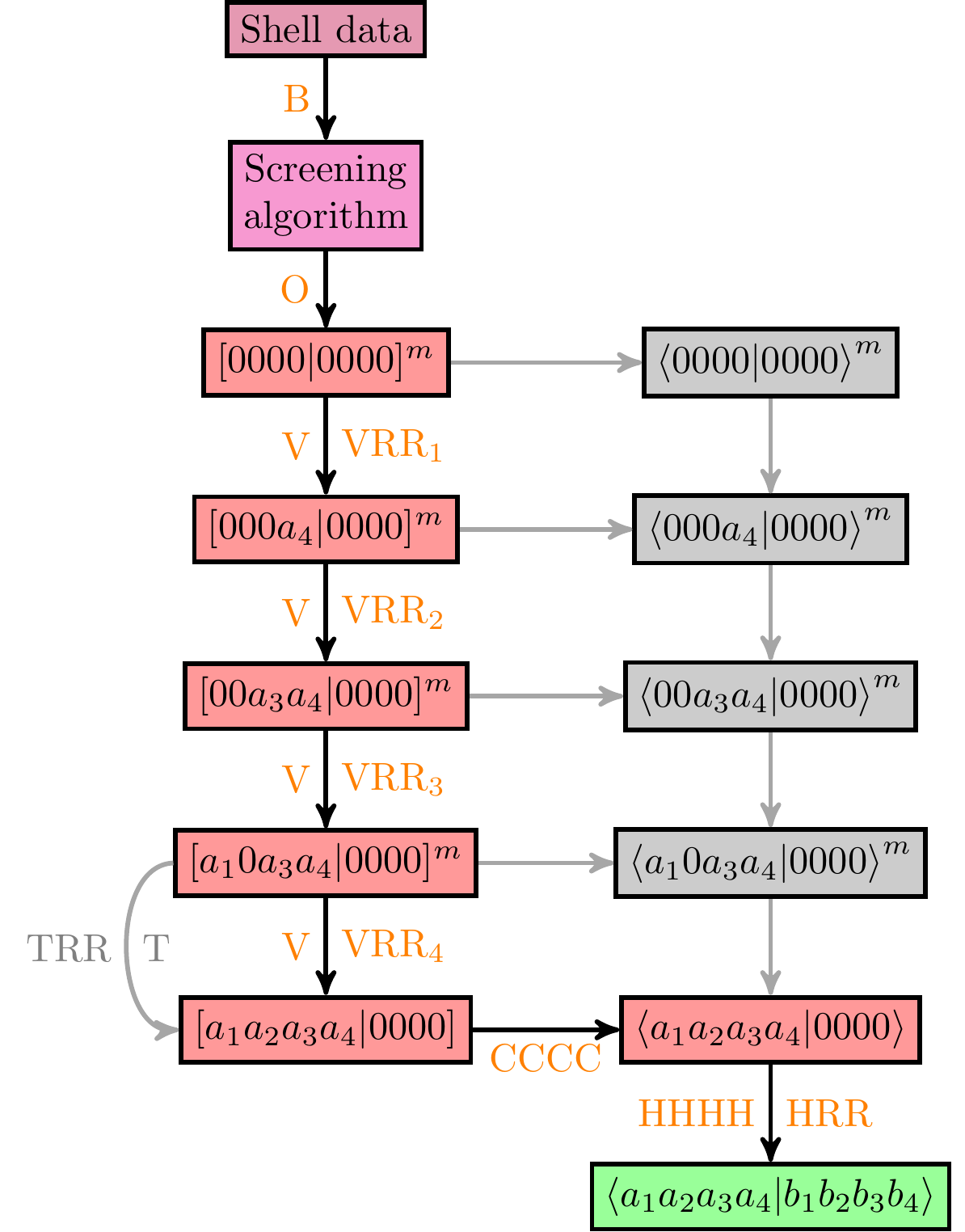}
\caption{
\label{fig:algo}
PRISM representation \cite{PRISM91} of the recursive algorithm used to compute a four-electron integral class $\braket{a_1 a_2 a_3 a_4}{b_1 b_2 b_3 b_4}$ over the trident operator $C_{12}G_{13}G_{14}$.
In this work, we consider the (orange) BOVVVVCCCCHHHH path.}
\end{figure}

\subsection{
\label{sec:cb}
Contracted bounds}
Contracted integral bounds are straightforward variations of primitive ones.
While contracting at the shell-pair level ($m=2$) only requires $\order{K^2}$ computational work, contracting at the shell-quartet, -sextet or -octet level would require $\order{K^4}$, $\order{K^6}$ or $\order{K^8}$ work, respectively.

Therefore, as sketched in Fig.~\ref{fig:scheme1}, we use a primitive bound for a first screening of the shell-pairs, then contracted bounds are used to screen shell-pairs, -quartets, -sextets and -octets.
Considering $K$-fold CGFs, the contraction step never exceeds $\mathcal{O}(K^{2})$ computational cost.
Bound factors such as
\begin{equation}
\label{eq:ct4}
	\expval{1}^{13,14} = \sum_{ij}^{K}{\sexpval{1}^{13,14}_{ij}},
\end{equation}
and its maximum within the basis set $\cmax{1}^{13,14}$, are computed within the shell-pair loop.
In Eq.~\eqref{eq:ct4}, $i$ and $j$ refer to the PGFs $\sket{a_{1}}_{i}$ and $\sket{b_{1}}_{j}$ in the contracted shells $\ket{a_{1}}$ and $\ket{b_{1}}$, respectively (see Eq.~\eqref{eq:CGTO-pair}).

The expressions of the contracted bounds are identical to the primitive bounds, with the only exception that the contracted factors $\expval{13}^{12,13,23}$, $\expval{123}^{12,13,23}$ and $\expval{134}^{12,13,14,34}$ are bound by
\begin{subequations}
\begin{align}
\label{eq:ct1}
		\expval{13}^{12,13,23} & \le \min \qty{ \expval{1}^{12,13} \expval{3}^{23},
		\expval{1}^{12} \expval{3}^{13,23} } \exp(-\check{Y}^{13}) ,
		 \\
\label{eq:ct2}
		\expval{123}^{12,13,23} & \le \expval{2} \min \qty{\expval{1}^{12,13}  \expval{3}^{23},
		\expval{1}^{12}  \expval{3}^{13,23} }\exp(-\check{Y}^{13,23}),
		 \\
\label{eq:ct3}
		 \expval{134}^{12,13,14,34} & \le  \min \qty{\expval{1}^{12,13} \expval{3}^{34} \expval{4}^{14},
		  \expval{1}^{12,14} \expval{3}^{13} \expval{4}^{34} } \exp(-\check{Y}^{13,14,34}),
\end{align}
\end{subequations}
where
\begin{equation}
	\check{Y}^{13,14,23,34}=  \Tr( \bm{\Db}_{0}^{{13},{14},{23},{34}} \cdot \bm{\check{Y}}^2),
\end{equation}
can be evaluated with the following expressions
\begin{subequations}
\begin{gather}
	\bm{\check{\delta}}_{u}^{{13},{14},{23},{34}}
	= \bm{\check{\zeta}} + \bm{\la}_{u} 
	= 
	\begin{pmatrix}
		\check{\zeta}_{1}	&	0		&	0		&	0		\\
		0		&	\check{\zeta}_{2}	&	0		&	0		\\
		0		&	0		&	\check{\zeta}_{3}	&	0		\\
		0		&	0		&	0		&	\check{\zeta}_{4}	\\
	\end{pmatrix}
	+
	\bm{\la}_{u},
	\\
	 \bm{\Db}_{u}^{{13},{14},{23},{34}} = \bm{\check{\zeta}} \cdot \left(\bm{\check{\delta}}_{u}^{{13},{14},{23},{34}}\right)^{-1} \cdot  \bm{\check{\zeta}},
	\\
	\bm{\check{Y}}^k = 
	\begin{pmatrix}
		0	&	\check{Y}_{12}^k	&	\check{Y}_{13}^k	&	\check{Y}_{14}^k	\\
		0	&	0			&	0			&	\check{Y}_{24}^k	\\
		0	&	0			&	0			&	\check{Y}_{34}^k	\\
		0	&	0			&	0			&	0			\\
	\end{pmatrix},
\end{gather}
\end{subequations}
where $\check{\zeta}_{i}$ is the smallest effective exponent $\sbzeta_{i}$ in the contracted shell-pair $\ket{a_ib_i}$, and
\begin{equation}
\label{eq:ct9}
	\check{Y}_{ij}= \max \qty{ 0, \abs{\frac{\bA_i \bB_i^+}{2}-\frac{\bA_j \bB_j^+}{2} } - \abs{ \frac{\bA_i \bB_i}{2}} - \abs{ \frac{\bA_j \bB_j}{2}} }
\end{equation} 
is the distance between two spheres of diameters $\bA_i \bB_i$ and $\bA_j \bB_j$ (where $\bA_i \bB_i^+ = \bA_i + \bB_i$).

\begin{table*}
\caption{
\label{tab:RRcount}
Number of intermediates required to compute various integral classes for two-, three- and four-electron operators.
The path generating the minimum number of intermediates is highlighted in bold.
The number of terms in the RRs and the associated incremental center are also reported.} 
\begin{ruledtabular}
\begin{tabular}{llllllcccc}
Integral			&	 type		&	operator				&	path		& number		& centers		&	\mc{4}{c}{integral class}	\\
																			\cline{7-10}
				&	 		&						&			& of terms		&	&	$\sexpval{p \ldots p}$	&	$\sexpval{d \ldots d}$	&	$\sexpval{f \ldots f}$	&	$\sexpval{g \ldots g}$	\\
\hline
two-electron		&	chain   	&	$G_{12}$				&	\bf VV		&	\bf (2,3)		&	($\bA_2$,$\bA_1$)					&	\bf 3		&	\bf 6	&	\bf 10	&	\bf 15		\\
				&		   	&						&	VT			&	(2,4)			&	($\bA_2$,$\bA_1$)					&	4		&	9	&	16	&	25		\\	
				&		   	&	$C_{12}$				&	\bf VV		&	\bf (4,6)		&	($\bA_2$,$\bA_1$)					&	\bf 4		&	\bf 13	&	\bf 25	&	\bf 48	\\
				&		   	&						&	VT			&	(4,4)			&	($\bA_2$,$\bA_1$)					&	7		&	19	&	37	&	61		\\
three-electron		&	chain   	&	$G_{13}G_{23}$		&	\bf VVV		&	\bf (2,3,4)		&	($\bA_3$,$\bA_2$,$\bA_1$)			&	\bf 5		&	\bf 13	&	\bf 26	&	\bf 45	\\
				&		   	&						&	VVT			&	(2,3,6)		&	($\bA_3$,$\bA_1$,$\bA_2$)			&	8		&	25	&	56	&	105		\\
				&		   	&	$C_{12}G_{23}$		&	VVV			&	(4,5,7)		&	($\bA_3$,$\bA_1$,$\bA_2$)			&	11		&	39	&	96	&	195		\\
				&		  	&						&	\bf VVV		&	\bf (4,6,6)		&	($\bA_3$,$\bA_2$,$\bA_1$)			&	\bf 10	&	\bf 39	&	\bf 96	&	\bf 196	\\
				&		  	&						&	VVT			&	(4,5,6)		&	($\bA_3$,$\bA_1$,$\bA_2$)			&	16		&	66	&	173	&	359		\\
				&		  	&						&	VVT			&	(4,6,6)		&	($\bA_3$,$\bA_2$,$\bA_1$)			&	15		&	65	&	171	&	357		\\
				&	cyclic	&	$C_{12}G_{13}G_{23}$	&	\bf VVV		&	\bf (4,6,8)		&	($\bA_3$,$\bA_2$,$\bA_1$)			&	\bf 12	&	\bf 46	&	\bf 119	&	\bf 250	\\
				&		   	&						&	VVT			&	(4,6,6)		&	($\bA_3$,$\bA_2$,$\bA_1$)			&	16		&	66	&	173	&	359		\\	
four-electron		&	chain   	&	$C_{12}G_{14}G_{23}$	&	VVVV		&	(4,5,7,8)		&	($\bA_4$,$\bA_3$,$\bA_2$,$\bA_1$)	&	21		&	108	&	344	&	847		\\						&		   	&						&	\bf VVVV		&	\bf (4,6,6,8)	&	($\bA_4$,$\bA_1$,$\bA_3$,$\bA_2$)	&	\bf 19	&	\bf 88	&	\bf 260	&	\bf 607	\\
				&		   	&						&	VVVT		&	(4,5,7,8)		&	($\bA_4$,$\bA_3$,$\bA_2$,$\bA_1$)	&	33		&	208	&	736	&	1,926	\\
				&		   	&						&	VVVT		&	(4,6,6,8)		&	($\bA_4$,$\bA_1$,$\bA_3$,$\bA_2$)	&	33		&	204	&	716	&	1,866	\\		
				&		   	&	$C_{12}G_{13}G_{34}$	&	VVVV		&	(4,6,6,9)		&	($\bA_4$,$\bA_3$,$\bA_2$,$\bA_1$)	&	22		&	113	&	360	&	888		\\
				&		   	&						&	\bf VVVV		&	\bf (4,6,8,7)	&	($\bA_4$,$\bA_3$,$\bA_1$,$\bA_2$)	&	\bf 20	&	\bf 98	&	\bf 302	&	\bf 726	\\
				&		   	&						&	VVVT		&	(4,6,6,8)		&	($\bA_4$,$\bA_3$,$\bA_2$,$\bA_1$)	&	33		&	204	&	716	&	1,866	\\
				&		   	&						&	VVVT		&	(4,6,8,8)		&	($\bA_4$,$\bA_3$,$\bA_1$,$\bA_2$)	&	34		&	214	&	756	&	1,976	\\
				&	trident   	&	$C_{12}G_{13} G_{14}$	&	VVVV		&	(4,6,6,9)		&	($\bA_4$,$\bA_3$,$\bA_2$,$\bA_1$)	&	22		&	113	&	360	&	888		\\
				&		   	&						&	\bf VVVV		&	\bf (4,6,8,7)	&	($\bA_4$,$\bA_3$,$\bA_1$,$\bA_2$)	&	\bf 20	&	\bf 98	&	\bf 302	&	\bf 726	\\
				&		   	&						&	VVVT		&	(4,6,6,8)		&	($\bA_4$,$\bA_3$,$\bA_2$,$\bA_1$)	&	33		&	204	&	716	&	1,866	\\
				&		   	&						&	VVVT		&	(4,6,8,8)		&	($\bA_4$,$\bA_3$,$\bA_1$,$\bA_2$)	&	34		&	214	&	756	&	1,976	\\
\end{tabular}
\end{ruledtabular}	
\end{table*}

\section{
\label{sec:RR}
Recurrence relations}
\subsection{
\label{sec:VRR}
Vertical recurrence relations}
Following Obara and Saika, \cite{OS1, OS2} vertical RRs (VRRs) are obtained by differentiation of Eq.~\eqref{eq:Fund0m} with respect to the centers coordinates. \cite{Ahlrichs06, 3ERI1}
For the integrals considered in this study, one can show that
\begin{equation}
\begin{split}
\label{eq:VRR}
	\sexpval{\cdots \ba_i^+ \cdots}^{m} 
	& = 
	\qty( \bZ_i\bA_i - \hD_i Y_0 ) \sexpval{\cdots \ba_i \cdots}^{m} 
	- \qty( \hD_i Y_1 - \hD_i Y_0 ) \sexpval{\cdots \ba_i \cdots}^{m+1}
	\\
	& + \sum_{j=1}^n \ba_j \Bigg\{ \qty( \frac{\delta_{ij}}{2\zeta_i} - \hD_{ij} Y_0 )  \sexpval{\cdots \ba_j^- \cdots}^{m} 
	 - \qty( \hD_{ij} Y_1 - \hD_{ij} Y_0 ) \sexpval{\cdots \ba_j^- \cdots}^{m+1} \Bigg\},
	\end{split}
\end{equation}
where $\delta_{ij}$ is the Kronecker delta, \cite{NISTbook}
\begin{align}
	\hD_i & = \frac{\nabla_{A_i}}{2\alpha_i},
	&
	\hD_{ij} & = \hD_i \hD_j,
\end{align}
and
\begin{subequations}
\begin{align}
	\hD_i Y_u & = \Tr(\bm{\Delta}_u \cdot \hD_i \bY^2),
	&
	(\hD_i \bY^2)_{kl} & = \kappa_{ikl} (\bY)_{kl},
	\\
	\hD_{ij} Y_u & = \Tr(\bm{\Delta}_u \cdot \hD_{ij} \bY^2),
	&
	(\hD_{ij} \bY^2)_{kl} & = \frac{\kappa_{ikl}\kappa_{jkl}}{2},
\end{align}
\end{subequations}
with
\begin{align}
	\eps_{ij} & = 
	\begin{cases}
		1,	&	\text{if $i \le j$},
		\\
		0,	&	\text{otherwise},
	\end{cases}
	&
	\kappa_{ijk} & = \frac{\eps_{ij} \delta_{ki} - \delta_{ij} \eps_{ki} }{\zeta_i}.
\end{align}
One can easily derive VRRs for other three- and four-electron operators following the simple rules given in Fig.~\ref{fig:tree}.
The number of terms for each of these VRRs is reported in Table \ref{tab:RRcount} for various two-, three- and four-electron operators.

Note that for a pure GG operator, we have $m = 0$ and $Y_1 = Y_0$.
Therefore, Eq.~\eqref{eq:VRR} reduces to a simpler expression:
\begin{equation}
\label{eq:VRR-pure}
	\sexpval{\cdots \ba_i^+ \cdots}
	= \qty( \bZ_i\bA_i - \hD_i Y_0 ) \sexpval{\cdots \ba_i \cdots} 
+ \sum_{j=1}^n \ba_j \qty( \frac{\delta_{ij}}{2\zeta_i} - \hD_{ij} Y_0 )  \sexpval{\cdots \ba_j^- \cdots}.
\end{equation}

\subsection{
\label{sec:TRR}
Transfer recurrence relations}
Transfer RRs (TRRs) redistribute angular momentum between centers referring to different electrons. \cite{3ERI1}
Using the translational invariance, one can derive 
\begin{equation}
\label{eq:TRR}
	\sexpval{\cdots \ba_i^+ \cdots}
	= \sum_{j=1}^n \frac{\ba_j}{2\zeta_i } \sexpval{\cdots \ba_j^- \cdots}
	- \sum_{j \neq i}^n \frac{\zeta_j}{\zeta_i } \sexpval{\cdots \ba_j^+ \cdots}
	- \frac{\sum_{j=1}^n \beta_j \bA_j \bB_j}{\zeta_i } \sexpval{\cdots \ba_j \cdots}.
\end{equation}
Note that Eq.~\eqref{eq:TRR} can only be used to build up angular momentum on the last center.
Moreover, to increase the momentum by one unit on this last center, one must increase the momentum by the same amount on all the other centers (as evidenced by the second term in the right-hand side of \eqref{eq:TRR}).
Therefore, the TRR is computationally expensive for three- and four-electron integrals due to the large number of centers (see below).
As mentioned by Ahlrichs, \cite{Ahlrichs06} the TRR can be beneficial for very high angular momentum two-electron integral classes.

\subsection{
\label{sec:HRR}
Horizontal recurrence relations}
The so-called horizontal RRs (HRRs) enable to shift momentum between centers over the same electronic coordinate: \cite{3ERI1}
\begin{equation}
\label{eq:HRR}
	\braket{\cdots \ba_i \cdots}{\cdots \bb_i^+ \cdots}
	= \braket{\cdots \ba_i^+ \cdots}{\cdots \bb_i \cdots}
	 + \bA_i \bB_i \braket{\cdots \ba_i \cdots}{\cdots \bb_i \cdots}.
\end{equation}
Note that HRRs can be applied to contracted integrals because they are independent of the contraction coefficients and exponents.

\begin{figure*}
	\includegraphics[height=0.38\textheight]{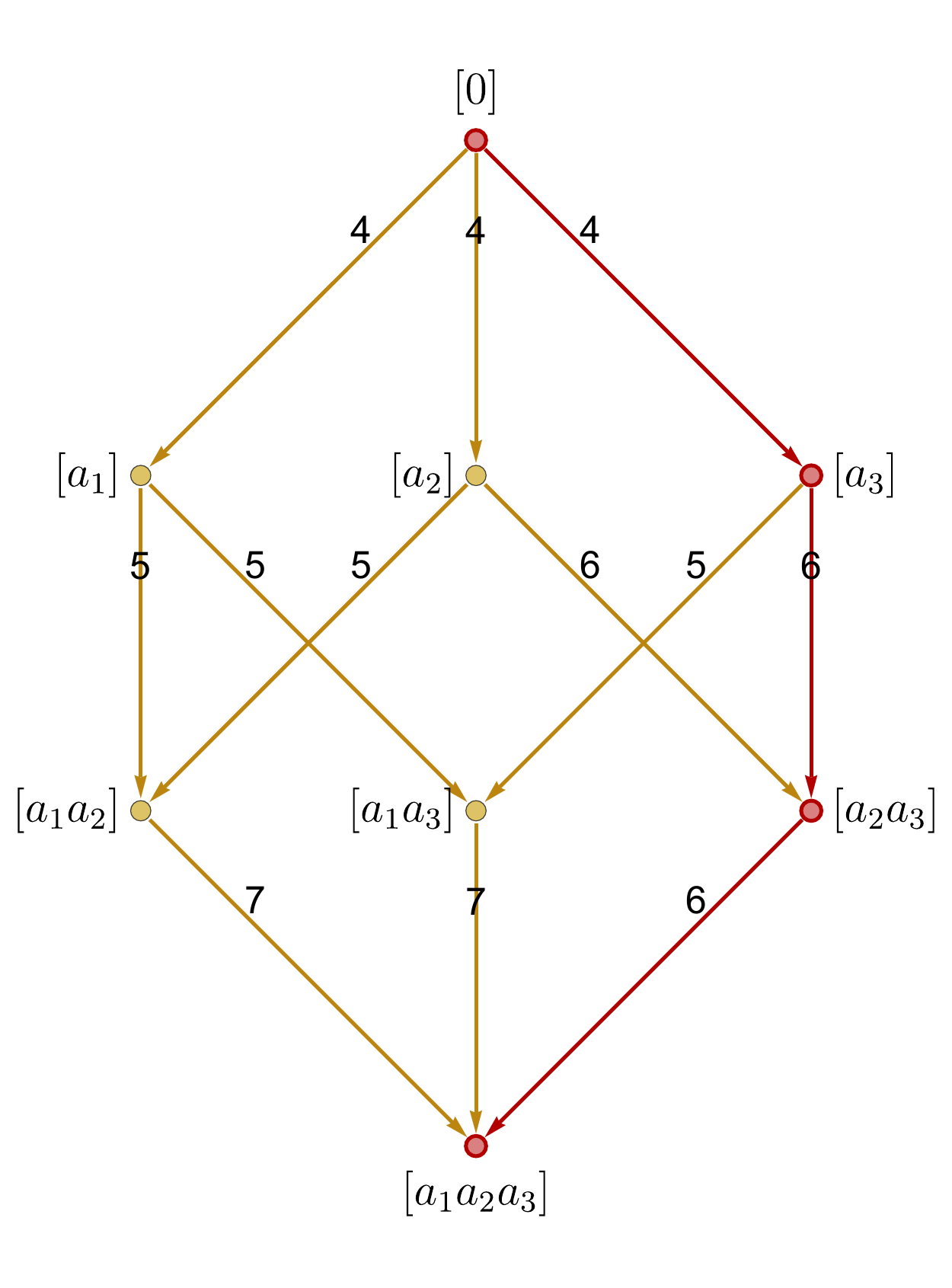}
	\includegraphics[height=0.38\textheight]{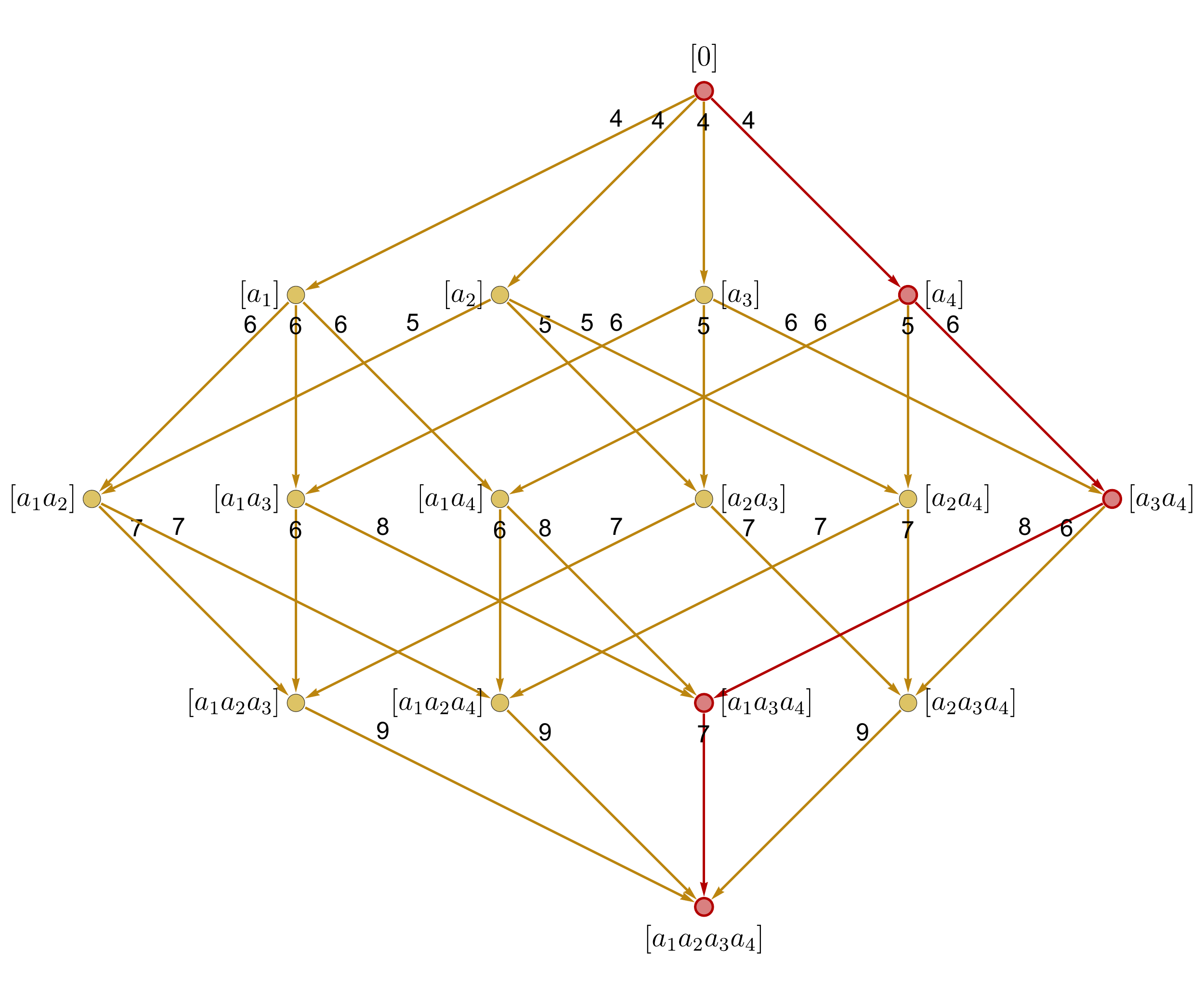}
\caption{
\label{fig:graph}
Graph representation of the VRRs for the 3-chain operator $C_{12}G_{23}$ (left) and trident operator $C_{12}G_{13} G_{14}$ (right).
The edge label gives the number of terms in the corresponding VRR.
The red path corresponds to the algorithm generating the smallest number of intermediates.}
\end{figure*}

\section{
\label{sec:algorithm}
Algorithm}
In this Section, we propose a recursive algorithm for the computation of a class of three- or four-electron integrals of arbitrary angular momentum.
The present recursive algorithm is based on a late-contraction scheme inspired by the Head-Gordon-Pople algorithm \cite{HGP} following a BOVVVVCCCCHHHH path.
The general skeleton of the algorithm is shown in Fig.~\ref{fig:algo} for the trident operator $C_{12}G_{13}G_{14}$.
We will use this example to illustrate each step.

Based on the shell data, the first step of the algorithm (step B) is to decide whether or not a given class of integrals is significant or negligible.
If the integral class is found to be significant by the screening algorithm presented in Sec.~\ref{sec:UB} and depicted in Fig.~\ref{fig:scheme1}, an initial set of FIs is computed (step O) via the formulae gathered in Sec.~\ref{sec:FI}.

Starting with these FIs, angular momentum is then built up over the different bra centers $\bA_1$, $\bA_2$, $\bA_3$ and $\bA_4$ using the VRRs derived in Sec.~\ref{sec:VRR}.
To minimize the computational cost, one has to think carefully how to perform this step.
Indeed, the cost depends on the order in which this increase in angular momentum is performed.
This is illustrated in Fig.~\ref{fig:graph}, where we have represented the various possible pathways for the 3-chain operator $C_{12}G_{23}$ (left) and the trident operator $C_{12}G_{13}G_{14}$ (right).
The red path corresponds to the path generating the least intermediates (i.e.~requiring the smallest number of classes in order to compute a given class).
Different paths are compared in Table \ref{tab:RRcount} for various two-, three- and four-electron operators, where we have reported the number of intermediates generated by each path for various integral classes.

Taking the 3-chain operator $C_{12}G_{23}$ as an example, one can see that, to compute a $\sexpval{ppp}$ class, it is more advantageous to build momentum over center $\bA_3$, then over centers $\bA_2$, and finally over center $\bA_1$ using VRRs with 4, 6 and 6 terms, respectively.
The alternative path corresponding to building momentum over $\bA_3$, $\bA_1$, and then $\bA_2$ with 4-, 5- and 7-term VRRs is slightly more expensive for a $\sexpval{ppp}$ class but becomes affordable for high angular momentum classes.
For both paths, using the TRR instead of the last VRR implies a large increase in the number of intermediates.

For the trident operator, we successively build angular momentum over $\bA_4$, $\bA_3$, $\bA_1$ and $\bA_2$ using VRRs with 4, 6, 8 and 7 terms.
The pathway using VRRs with 4, 6, 6, and 9 terms is more expensive due to the large number of terms of the VRR building up momentum over the last center.
Again, using the TRR instead of the last VRR significantly increases the number of intermediates.

The path involving the minimal number of intermediates is given in Table \ref{tab:RRcount} for various two-, three- and four-electron operators. 
It is interesting to point out that it is never beneficial to use the TRR derived in Eq.~\eqref{eq:TRR} (see Sec.~\ref{sec:TRR}).

One can easily show that, for operators involving the Coulomb operator, the number of intermediates required to compute a $n$-electron integral class $\sexpval{a \ldots a}$ increases as $\order{a^{n+1}}$ for the VRR-only paths (see Table \ref{tab:RRcount}).
This number is reduced to $\order{a^{n}}$ if one uses the TRR to build up angular momentum on the last center. 
However, the prefactor is much larger and the crossover happens for extremely high angular momentum for three- and four-electron integrals.
For ``pure''  GG operators, such as $G_{12}$ or $G_{13}G_{23}$, the number of intermediates required to compute a class $\sexpval{a \ldots a}$ increases as $\order{a^n}$ for any type of paths.

Finally, we note that the optimal path for the trident $C_{12}G_{13}G_{14}$ and the 4-chain $C_{12}G_{13}G_{34}$ is similar, thanks to their similar structure.
Indeed, these two operators can be seen as two ``linked'' GGs ($G_{13}G_{14}$ or $G_{13}G_{34}$) interacting with the Coulomb operator $C_{12}$ (see Fig.~\ref{fig:tree}), while the other 4-chain operator $C_{12}G_{14}G_{23}$ can be seen as two ``unlinked'' GGs ($G_{14}$ and $G_{23}$) interacting with the Coulomb operator.

When angular momentum has been built over all the bra centers, following the HGP algorithm, \cite{HGP} we contract $\sbraket{\ba_1 \ba_2 \ba_3  \ba_4}{\bo \bo \bo \bo}$ to form $\braket{\ba_1 \ba_2 \ba_3  \ba_4}{\bo \bo \bo \bo}$ (step CCCC).
We can perform the contraction at this point because all of the subsequent RRs are independent of the contraction coefficients and exponents.
More details about this contraction step can be found in Ref.~\onlinecite{Gill94b}.

The last step of the algorithm (step HHHH) shifts momentum from the bra center $\bA_1$, $\bA_2$, $\bA_3$ and $\bA_4$ to the ket centers  $\bB_1$, $\bB_2$, $\bB_3$ and $\bB_4$ using the two-term HRRs given by Eq.~\eqref{eq:HRR} in Sec.~\ref{sec:HRR}.

\section{Concluding remarks}
We have presented the three main ingredients to compute three- and four-electron integrals involving GGs.
Firstly, a straightforward method to compute the FIs is given.
Secondly, scaling-consistent UBs are reported, as they allow to evaluate only the $\order{N^{2}}$ significant integrals in a large system. 
Finally, the significant integrals are computed via a recursive scheme based on vertical and horizontal RRs, which can be viewed as an extension of the PRISM late-contraction path to three- and four-electron integrals. 

We believe our approach represents a major step towards an accurate and efficient computational scheme for three- and four-electron integrals.
It also paves the way to contraction-effective methods for these types of integrals. 
In particular, an early-contraction scheme would have significant computational benefits.

\begin{acknowledgments}
The authors thank Peter Gill for many stimulating discussions.
P.F.L.~thanks the Australian Research Council for a Discovery Project grant (DP140104071) and the NCI National Facility for generous grants of supercomputer time. 
Financial support from the \textit{Centre national de la recherche scientifique} is also acknowledged.
\end{acknowledgments}

%

\end{document}